\documentclass[a4paper,12pt]{article}
\usepackage[dvips]{graphics}
\usepackage{rotating,epsfig}
\newcommand{\minisize}  {16.0cm}

\newcommand{\inmath}[1] {\ifmmode#1\else$#1$\fi}
\newcommand{\definmath}[2] {\def#1{\ifmmode#2\else$#2$\fi}}

\def\D0{\mbox{$D^{0}$\ }}

\definmath{\roots} {\sqrt{s}}
\definmath{\Ecm} {E_{\mathrm{cm}}}
\definmath{\Ebeam}  {E_{\mathrm{b}}}
\definmath{\xE} {x_E}
\definmath{\as} {\alpha_s}
\definmath{\Evis}   {E_{\mathrm{vis}}}

\definmath{\PZz} {\mathrm{Z}^{0}}      
\definmath{\PWpm} {\mathrm{W}^{\pm}}      
\definmath{\Plp} {\ell^{+}}        
\definmath{\Plm} {\ell^{-}}        
\definmath{\Plpm}   {\ell^{\pm}}         
\definmath{\Pgtp} {\tau^{+}}        
\definmath{\Pgtm} {\tau^{-}}        
\definmath{\Pgtpm}   {\tau^{\pm}}         
\definmath{\Pgn}  {\nu}          
\definmath{\Pagn} {\overline{\nu}}     
\definmath{\Pq}      {\mathrm{q}}
\definmath{\Paq}  {\overline{\mathrm{q}}}
\definmath{\Puds}      {\mathrm{uds}}
\definmath{\Pu}      {\mathrm{u}}
\definmath{\Pau}  {\overline{\mathrm{u}}}
\definmath{\Pd}      {\mathrm{d}}
\definmath{\Pad}  {\overline{\mathrm{d}}}
\definmath{\Ps}      {\mathrm{s}}
\definmath{\Pas}  {\overline{\mathrm{s}}}
\definmath{\Pc}      {\mathrm{c}}
\definmath{\Pac}  {\overline{\mathrm{c}}}
\definmath{\Pb}      {\mathrm{b}}
\definmath{\Pab}  {\overline{\mathrm{b}}}
\definmath{\Pg}      {\mathrm{g}}
\newcommand{\ov}[1]{$\overline{\rm #1}$}
\newcommand{\bit}{\begin{itemize}}
\newcommand{\eit}{\end{itemize}}
\newcommand{\ben}{\begin{enumerate}}
\newcommand{\een}{\end{enumerate}}
\newcommand{\bde}{\begin{description}}
\newcommand{\ede}{\end{description}}
\newcommand{\bce}{\begin{center}}
\newcommand{\ece}{\end{center}}
\newcommand{\beq}{\begin{equation}}
\newcommand{\eeq}{\end{equation}}
\newcommand{\Rule}{\rule[-.7ex]{0ex}{4.ex}}

\def\x0{~$X_{0}$}
\def\GeV{{\rm GeV}}

\def\sigmaE{\sigma_{E}}
\def\Dp{\Delta\phi}
\def\DT{\Delta\Theta}
\def\Dcut{\Delta_{\rm cut}}
\def\Scut{S_{{\rm cut}}}

\def\Dgq{D_{\gamma/{\rm q}}}

\def\Q2{Q^{2}}
\def\Zo{$\rm Z^0$}

\def\ppio{\rm \pi^{0}}
\def\pic{\pi^{\pm}}
\def\ee{$\rm {e^+e^-}$}

\begin{document}
\begin{titlepage}
\begin{center}{\large   EUROPEAN LABORATORY FOR PARTICLE PHYSICS
}\end{center}\bigskip
\begin{flushright}
       CERN-PPE/97-086   \\ 17 July 1997
\end{flushright}
\bigskip\bigskip\bigskip\bigskip\bigskip
\begin{center}{\huge\bf   
Measurement of the quark to photon \\ 
\vspace{0.2cm}
fragmentation function through the \\ 
\vspace{0.2cm}
inclusive production of prompt \\ 
\vspace{0.22cm}
photons in hadronic $\rm \bf Z^0$\thinspace decays
}\end{center}\bigskip\bigskip
\begin{center}{\LARGE The OPAL Collaboration
}\end{center}\bigskip\bigskip
\bigskip\begin{center}{\large  Abstract}\end{center}
  The inclusive production of prompt photons with energy above 10 GeV
  is measured using the  OPAL detector in hadronic \Zo ~decays at LEP.  In
  contrast to previous measurements, the prompt photons were not
  required to be isolated.  The production rate and energy spectrum
  are found to be in agreement with  QCD predictions for the
  quark-to-photon fragmentation function.  
\bigskip\bigskip\bigskip\bigskip
\bigskip\bigskip
\begin{center}{\large
(To be submitted to Zeitschrift f\"{u}r Physik C)
}\end{center}
\end{titlepage}
\begin{center}{\Large        The OPAL Collaboration
}\end{center}\bigskip
\begin{center}{
K.\thinspace Ackerstaff$^{  8}$,
G.\thinspace Alexander$^{ 23}$,
J.\thinspace Allison$^{ 16}$,
N.\thinspace Altekamp$^{  5}$,
K.J.\thinspace Anderson$^{  9}$,
S.\thinspace Anderson$^{ 12}$,
S.\thinspace Arcelli$^{  2}$,
S.\thinspace Asai$^{ 24}$,
D.\thinspace Axen$^{ 29}$,
G.\thinspace Azuelos$^{ 18,  a}$
% ...
}\end{center}\bigskip
\bigskip
% ...
$^{  2}$Dipartimento di Fisica dell' Universit\`a di Bologna and INFN,
I-40126 Bologna, Italy
\newline
% ...
$^{  5}$Cavendish Laboratory, Cambridge CB3 0HE, UK
\newline
% ...
$^{  8}$CERN, European Organisation for Particle Physics,
CH-1211 Geneva 23, Switzerland
\newline
$^{  9}$Enrico Fermi Institute and Department of Physics,
University of Chicago, Chicago IL 60637, USA
\newline
% ...
$^{ 12}$Indiana University, Department of Physics,
Swain Hall West 117, Bloomington IN 47405, USA
\newline
% ...
$^{ 16}$Department of Physics, Schuster Laboratory, The University,
Manchester M13 9PL, UK
\newline
% ...
$^{ 18}$Laboratoire de Physique Nucl\'eaire, Universit\'e de Montr\'eal,
Montr\'eal, Quebec H3C 3J7, Canada
\newline
% ...
$^{ 23}$Department of Physics and Astronomy, Tel Aviv University,
Tel Aviv 69978, Israel
\newline
$^{ 24}$International Centre for Elementary Particle Physics and
Department of Physics, University of Tokyo, Tokyo 113, and
Kobe University, Kobe 657, Japan
\newline
% ...
$^{ 29}$University of British Columbia, Department of Physics,
Vancouver BC V6T 1Z1, Canada
\newline
% ...
$^{  a}$ and at TRIUMF, Vancouver, Canada V6T 2A3
\newline
% ...
\newline
\newpage
\section{Introduction}
    
   We gain a greater understanding of the 
properties of the elementary building blocks of matter and their 
interactions 
by studying the properties of hadrons, leptons and photons produced 
as a result of a primary interaction. The properties of hadrons 
and charged leptons produced in \ee ~collisions 
have been studied in great detail at different  centre-of-mass energies.   
In the case of photons radiated off quarks, prompt photons,  in 
hadronic \ee  ~collisions,  
much less information is available 
due to 
the difficulty of separating these photons from those produced in the 
 decays of other particles~\cite{CELLO}-\cite{VENUS}. Both the shape and normalisation of the 
inclusive prompt photon energy spectrum in \ee 
~collisions are predicted, through the calculation of the 
quark-to-photon fragmentation function, 
by leading-order perturbative QCD 
 \cite{EWitten,CHL-L}. 
This asymptotic  prediction  has been parametrised
 in \cite{DUKEOWENS}. 
Non-perturbative effects can be included in the calculation 
through the vector-meson dominance ansatz as in \cite{GRV,Fontannaz}, 
where boundary terms missing in \cite{DUKEOWENS} were also accounted for.
The higher-order terms were calculated, and seen to be small 
at the energies close to the \Zo ~peak. Direct experimental study  
of these predictions is important in providing insight into the
non-perturbative and higher order 
effects in the radiation of photons from quarks. 
This will also make theoretical predictions of photon production
in other processes, such as those occuring at pp 
and p\ov{p}
colliders, more reliable, 
thus improving sensitivity to possible new phenomena.

   At LEP the first measurement of prompt 
photon production in hadronic \Zo ~decays 
was made by the OPAL Collaboration \cite{OPAL1}
for photons isolated from other particles in the event,  as 
suggested in \cite{Peter}. The production of isolated prompt photons
was studied in great detail by all LEP experiments \cite{ALEPH}-\cite{OPAL}.
Following the suggestion of \cite{Glover}, the ALEPH Collaboration extracted 
the quark-to-photon fragmentation function from the study of non-isolated 
photons in jets containing a photon carrying more than 70\% of the jet 
energy~\cite{ALEPH_frfun}.

   Here we present a  measurement of the inclusive prompt 
photon energy spectrum in hadronic \Zo ~decays at LEP. This method of
studying the quark-to-photon fragmentation function was suggested in 
\cite{Fontannaz,Kunszt}. 
To separate prompt photons from the photons from decays of other particles 
we use the following method.   
We selected clusters in the
electromagnetic calorimeter not associated with charged tracks.
A set of cuts were applied to reduce the 
background in the sample. The distribution of a variable characterising
the transverse shape of the clusters in data was then fitted with a
linear combination of the distributions for photons and for background 
to determine the fraction of prompt 
photons in the selected sample. The result was then corrected for 
the selection efficiencies, detector effects and initial state radiation. 
   In the following sections we describe the OPAL detector, the 
event and electromagnetic cluster selection (sec.~\ref{sec_met}) and the 
determination of the number of photons in the selected sample 
(sec.~\ref{sec_fdat}). 
The efficiency and acceptance corrections 
are described (sec.~\ref{sec_cor}) followed by the study of systematic effects 
(sec.~\ref{sec_sys}). Finally the measured prompt photon energy
spectrum is presented and discussed (sec.~\ref{sec_result}).

\section{The OPAL detector}
\label{sec_det}

\setcounter{footnote}{0}

The OPAL detector operates at the LEP \ee ~collider at CERN.  A detailed
description of the detector can be found in \cite{OPALNIM}. 
For this study, the most important 
components of OPAL were the central detector   
and the barrel electromagnetic calorimeter with its presampling detector. 
The central detector, measuring the momenta of 
charged particles, consists of a system of cylindrical tracking 
chambers surrounded by a solenoidal coil which produces
a uniform axial  magnetic field of 0.435~T along the beam 
axis\footnote{In the OPAL coordinate system the $x$ axis points
towards the centre of the LEP ring, the $y$ axis points upwards and
the $z$ axis points in the direction of the electron beam.  The
polar angle $\theta$ and the azimuthal angle $\rm \phi$ are defined
with respect to the  $z$ and $x$-axes, respectively, 
while $r$ is the distance from the
$z$-axis.}. The detection efficiency for charged particles is almost 
100\% within the polar angle range $|\cos\theta|<0.95$.  

The electromagnetic calorimeters completely cover the azimuthal 
range for polar angles satisfying $|\cos\theta|<0.98$ providing 
excellent hermeticity. 
    The barrel electromagnetic calorimeter covers 
the polar angle range $|\cos\theta|<0.82$. It 
consists of 9440 lead glass blocks, each 24.6 radiation lengths deep, 
almost pointing towards the interaction region.  Each block subtends an 
angular region of approximately $40\times40~\rm{mrad}^{2}$. Half of the 
block width corresponds to 1.9 Moli\'ere radii. Deposits 
of energy in adjacent blocks are grouped together to form clusters
of electromagnetic energy.  The intrinsic energy resolution 
of $\rm {\sigmaE/E=0.2\%\oplus6.3\%/\sqrt{E}}$
is substantially degraded (by a factor $\simeq 2$)
due to the presence of  two radiation lengths of 
material in front of the lead glass. For the    
intermediate region, $0.72<|\cos\theta|<0.82$, the amount of material 
increases up to eight radiation lengths causing further degradation in 
the energy resolution. 
The two endcap calorimeters, each made of 
1132 lead glass blocks, 22 radiation lengths deep, cover the region of 
$0.81<|\cos\theta|<0.98$. 
In this study the measurement of inclusive photon production 
is restricted to the barrel part of the detector.    
Most of the electromagnetic showers start before the calorimeter and
their position at the entrance of the calorimeter
is measured by a barrel electromagnetic presampler 
made of limited streamer mode chambers. 
The presampler covers the polar angle range $|\cos\theta|<0.81$ and its
angular resolution for photons is approximately $2~\rm{mrad}$.
\section{The selection of events and electromagnetic clusters}
\label{sec_met}

   Our study was based on a sample of 2.5 million hadronic \Zo ~decays  
selected as described in \cite{Hadsel_paper} from the data accumulated with 
the OPAL detector at LEP in 1992, 1993 and 1994 at an 
\ee ~centre-of-mass energy
 of $\rm{91.2~GeV}$. We did not use the off-peak 
data to avoid additional complications in dealing with data collected at 
different \ee ~centre-of-mass
energies.    
  We required that the central detector and the calorimeters were 
fully operational.  Temporary,  local inefficiencies in the presampler
chambers were monitored and taken into account.

  To study the properties of the background we used
 Monte Carlo events produced with the parton shower 
generators JETSET~7.4 \cite{JETSET} 
(3.9 million events) and HERWIG~5.8 
\cite{HERWIG}(1.1 million events) with generator parameters given in 
\cite{JETSETUNE}. We also used samples of events with only single 
photon or a $\ppio$ meson present in  the detector.   
The Monte Carlo (MC) samples  
were passed through the full 
simulation of the OPAL detector \cite{GOPAL} and subjected to the same 
reconstruction and analysis procedure as the data.

   The difficulty in the measurement of prompt photon production 
lay in the separation of the signal from background. The QCD shower models 
predicted a signal-to-background ratio of approximately 1/200. 
Background clusters, with no charged track associated with them, 
are dominated by photons from decays of 
hadrons, particularly 
$\ppio\rightarrow\gamma\gamma$ ($\simeq~$57\%) and
$\eta\rightarrow\gamma\gamma$ ($\simeq$~10\%). 
The other sources of 
background, like the interaction of neutral hadrons such as 
$\rm K^{0}_{L}$'s or neutrons in the material of the calorimeter contribute
at the level of few per cent each.  
More than one particle can also contribute to a cluster. 
 The transverse profile of the calorimeter cluster 
can be used to differentiate between clusters coming from different 
sources. The hardest to remove are clusters 
produced by 
$\ppio\rightarrow\gamma\gamma$ and
$\eta\rightarrow\gamma\gamma$ 
meson decays 
because they can be very similar, especially for higher cluster energies, 
to those produced by one photon. 
An irreducible, but very well predicted, background comes 
from the initial state radiation (ISR) photons radiated by the beam particles
before they interacted. 

   The tracks and calorimeter
 clusters were selected as described in \cite{Hadsel_paper}. 
 In addition we required a cluster energy to be larger than $10~\GeV$ 
and cluster polar angle such 
that $|\cos\theta|<0.72$.  We then applied three cuts motivated by 
studies with simulated events. 

\begin{figure}[p]
\begin{minipage}{\minisize}
\bce
\mbox{
\epsfig{file=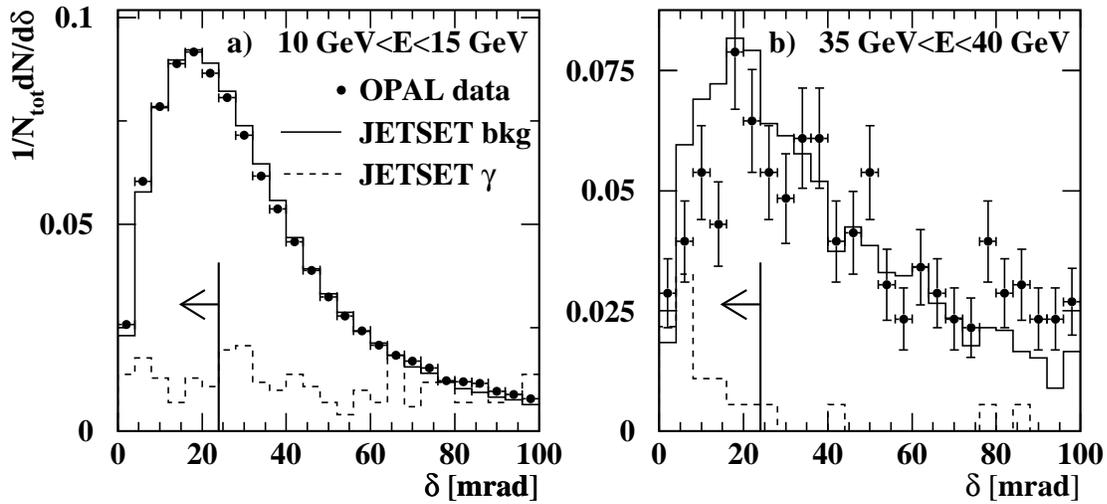,width=\minisize}
}
\vspace{-7.5cm}
\caption{The angle, $\delta$, between the calorimeter cluster and the 
nearest track for hadronic events: for the data, background clusters in 
the JETSET simulated event sample and prompt photons in the JETSET sample, 
for different 
cluster energies: $10<E<15~\GeV$ a), and  
$35<E<40~\GeV$ b). 
Distributions are normalised such that the total yield 
for each curve is unity. Most of the photon signal is off scale, 
with $\delta>\rm 100~mrad$.
The regions removed by the cut are shown by arrows.}
\label{fig_dmctr}
\ece
\end{minipage}
\end{figure}

\begin{figure}[p]
\begin{minipage}{\minisize}
\bce
\mbox{
\epsfig{file=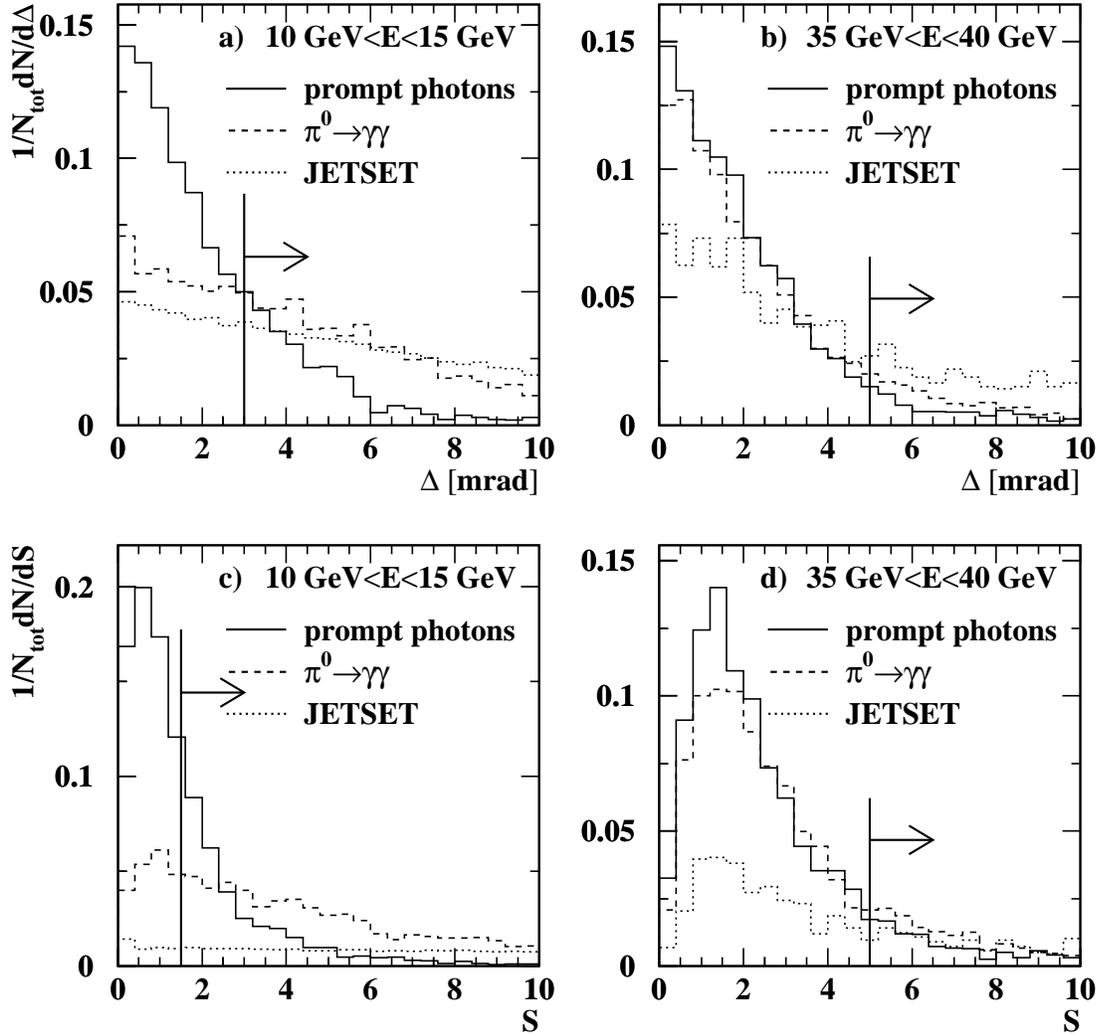,width=\minisize         }
}
\caption{The distributions of the variables $\Delta$ (a and b), and 
S (c and d) for clusters from
simulation of single isolated photons and $\ppio$'s and 
for all background clusters in JETSET events (after cut 1),  
for different 
cluster energies:  $ 10<E<15~\GeV$ a) and c), and 
$35<E<40~\GeV$ b) and d). 
Distributions are normalised such that the total yield 
for each curve is unity.
The regions removed by the cuts are shown by arrows.}
\label{fig_minds}
\ece
\end{minipage}
\end{figure}

\ben
\item[Cut 1.] There was required to be no charged track associated with the
 cluster. Tracks were extrapolated to the calorimeter surface. 
      A track was associated with a calorimeter cluster if
      it  extrapolates to 
      the calorimeter within 24~mrad (approximately
       half of a lead glass block
       width) of the centre of gravity of a cluster. In 
      Figure~\ref{fig_dmctr}a and~\ref{fig_dmctr}b, the normalised  
      distributions of the angle 
      between the calorimeter cluster and the nearest track are shown for 
      background clusters and prompt photons in the JETSET model, for  
      small and large cluster energies. 
      For the higher
      energy clusters a contribution from electrons and positrons
      from conversions of prompt photons
      in the beam pipe and central detector is seen  
      for small angles $\delta$.  
      For lower energy clusters this effect is diluted both by the  
      greater separation between the extrapolation of the track 
      to the calorimeter surface and the centre of gravity 
      of the calorimeter cluster, and  by the 
      presence of other tracks in the proximity of the prompt
      photon.
\item[Cut 2.] We required the presence of a
         presampler cluster within $24~\rm{mrad}$  
       of the centre of gravity of the calorimeter cluster. 
       The differences in azimuthal $|\Dp|$
       and polar $|\DT|$ angles
        between 
        the positions of the calorimeter cluster and the presampler cluster 
       were required to satisfy  
      $\Delta=\rm min(|\Dp|,|\DT|)<\Dcut$.
      The distributions of $\Delta$ for single, isolated photons and 
      $\ppio$'s in the detector
      as well as all background clusters in JETSET events are shown in 
      Fig.~\ref{fig_minds}a and \ref{fig_minds}b. 
      The value of $\Delta$ tended to be larger for clusters produced by two 
      overlapping photons, e.g. from 
      a $\ppio\rightarrow\gamma\gamma$
      than for clusters produced by a single photon,  
      because the presampler measures the cluster position at an 
      early stage of development of the electromagnetic shower.

\item[Cut 3.] The transverse profile of the cluster in the calorimeter
      was required to be compatible with 
      that produced by an isolated photon. 
      It was first assumed that the cluster was produced by a photon. 
      The impact point of the photon was varied until the best description 
      of the observed lateral shower profile by a reference profile
       was found. The reference profile was obtained by the
        parametrisation of the results of MC simulation of the 
        isolated photon
        in the detector. The fast algorithm
       described in section~4.1 of~\cite{taupol} was used. 
      The resulting variable, $S$, is proportional to the 
      $\chi^{2}$ for matching 
      the measured and predicted energy sharing between 
      the calorimeter blocks. 
      The distributions of $S$ for single photons, single $\ppio$'s 
      and background clusters in JETSET events are shown in 
      Fig.~\ref{fig_minds}c and \ref{fig_minds}d. 
\een
 The energy dependent 
      values of $\Dcut$ and $\Scut$ were chosen to 
      optimise the separation of signal from background and are shown in 
      Tab.~\ref{tab_cor}.

    In total  23106 clusters passed the selection procedure. 
The signal-to-background ratio, estimated from simulation, 
was improved from $1/200$ to $1/130$ after cut~1, 
then to $1/50$ after cut~2 and finally to $1/6$ after cut~3 with respectively 
92\%, 51\% and 42\% of the signal retained.   
According to the simulation the background clusters passing the selection
 were produced 
mostly by $\ppio\rightarrow\gamma\gamma$
($\simeq$79\%) and  
$\eta\rightarrow\gamma\gamma$ 
($\simeq$11\%) decays.

   The cuts lead to a strong reduction of the background while prompt photons
were much less affected. The fraction of photons rejected by the cuts 
can be corrected for and its knowledge will affect the systematic 
uncertainty of the measurement as detailed in Sections~\ref{sec_cor}
 and~\ref{sec_sys}. 
     The efficiency  of cuts~2 and~3 for photon clusters well separated
 from other particles in the event was determined directly from the 
data using a sample of photons in radiative lepton pair events 
\ee$\rightarrow \rm{\ell^{+}\ell^{-}\gamma}$ ($\ell=$~e,~$\rm \mu$). 
In addition, in hadronic events, further losses of prompt photons 
occur when other particles hit the calorimeter close to the photon. 
This can result in a cluster being associated to a track or
being sufficiently distorted 
to fail the selection criteria. A small fraction 
($\simeq6\%$) of prompt photons also converted 
in the beam pipe or central detector. 
The correction for these effects was estimated using Monte Carlo as  
detailed in sec.~\ref{sec_cor}. 
     In Figure~\ref{fig_dmctr}
we compare the data and MC distributions for the angle $\delta$ between 
the cluster and track closest to it. Differences between data and Monte Carlo 
are concentrated in the region of 
small angles, 
below our cut value of 24~mrad, a region containing only 
a small fraction of prompt photons.   
In Figure~\ref{fig_dmcds} we present data and MC distributions for  
the variables $S$ and $\Delta$ used in cuts 2 and 3. 
We compare distributions for photons from radiative lepton events  
 with MC for single, isolated photons (a and c). We also show    
 distributions for data and MC clusters from the  
 $\tau^{+}\tau^{-}$ events 
($\rm {\tau^{\pm}\rightarrow\rho^{\pm}\nu_{\tau}}$ 
and $\rm {\rho^{\pm}\rightarrow\pic\ppio}$, section~4.2 of \cite{taupol})
(b and d). 
The differences between data and MC are small and concentrated mostly in the 
region below our cut values, so they will cause only small systematic effects.
\section{The determination of the number of photons in the selected
sample} 
\label{sec_fdat}
 
We determined the fraction of photons in the sample  remaining after 
cuts 1-3 above  
using the cluster shape fit variable $C$  
used in the previous OPAL studies of photon production 
\cite{OPAL1,OPAL}. The fit algorithm applied 
was 
more sophisticated than that used in the 
the calculation of the variable $S$.
The variable $C$ had a better background rejection power than
$S$. Due to the similar fit 
algorithms and shower parametrisations the 
$C$ and $S$ variables are correlated, although not fully. 
The definition of $C$ is
\begin{equation}
\label{eq_C}
C= \frac{1}{N_{b}}\sum_{i}\frac{(E_{i}^{\rm pred}-E_{i}^{\rm obs})^{2}}
      {(\sigma_{i}^{\rm pred})^{2}+(\sigma_{i}^{\rm obs})^{2}},
\end{equation}
where: 
      $E_{i}^{\rm obs}$ is the energy observed in calorimeter block number $i$;
      $E_{i}^{\rm pred}$ is the
          predicted energy in calorimeter block number $i$;
      $\sigma_{i}^{\rm pred}$ and $\sigma_{i}^{\rm obs}$  are the 
       energy dependent errors 
      on $E_{i}^{\rm pred}$ and $E_{i}^{\rm obs}$, respectively; and
      $N_{b}$ is the number of blocks in the cluster.
     $E_{i}^{\rm pred}$ was taken from the best fit 
of the shower profile parametrisation, assuming that the cluster was
produced by a isolated photon, 
to the observed energy sharing between the calorimeter blocks.  
The reference profiles varied as a function of $\cos\theta$ because of the 
varying amount of material in front of the calorimeter.  

 We fitted the distribution of $C$  
in the data  with a linear combination of 
MC distributions for photons and background for clusters passing
the same selection criteria as the data:
\begin{equation}
\label{eq_fit}
           C_{\rm fit}=fC_{\gamma}+(1-f)C_{\rm bkg}.
\end{equation}

The fraction $f$ of photons in the selected sample was the fit parameter.
     For the distribution of the background, 
$C_{\rm bkg}$, clusters from JETSET hadronic \Zo ~events were used, where
initial state radiation and prompt photons were removed from the sample. 
     For the distribution of the photons, $C_{\gamma}$, we used 
a simulated sample of isolated photons.
     The fit for $f$  to the $C$ variable distribution 
in the data was performed separately in seven bins of cluster energy as 
shown in first column of Tables~\ref{tab_cor} and~\ref{tab_last}. 
The $C_{\gamma}$ and  $C_{\rm bkg}$ distributions had
only a small dependence on cluster energy within a given bin. 
A binned maximum likelihood method \cite{RBarlow} was used to  
fit the $C$ variable distribution between 0 and 5. 

\begin{figure}[p]
\begin{minipage}{\minisize}
\bce
\mbox{
\epsfig{file=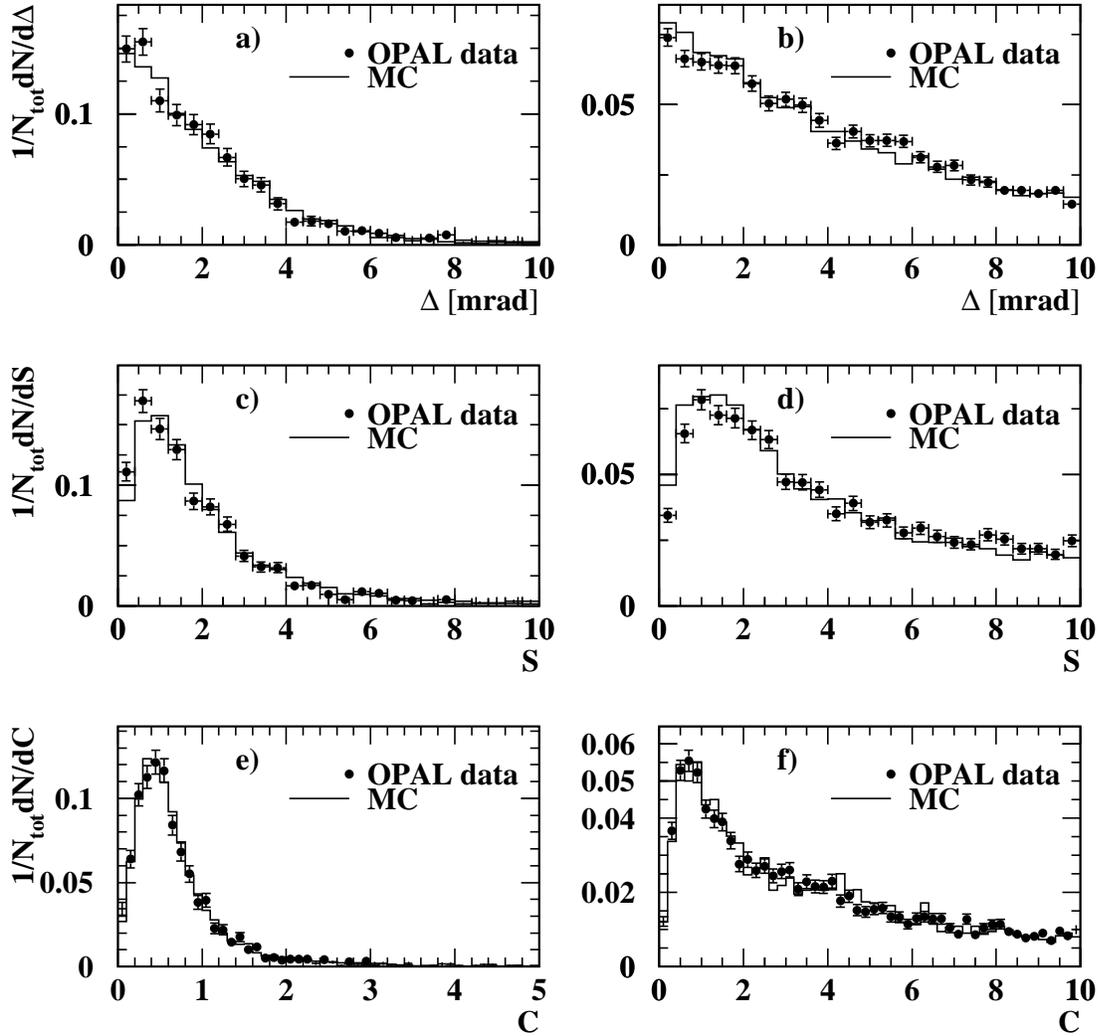,width=\minisize}
}
\caption{The distributions of the variables $\Delta$ (a and b), $S$ 
(c and d) and $C$ (e and f) 
for photons in radiative lepton pair data events and simulated 
single, isolated photons (a, c and e); and $\ppio$'s in 
$\tau^{\pm}\rightarrow\rho^{\pm}\nu_{\tau}$, 
$\rho^{\pm}\rightarrow\pic\ppio$ decays (b, d and f).
Distributions are normalised such that the total yield 
for each curve is unity}
\label{fig_dmcds}
\ece
\end{minipage}
\end{figure}

Since we used simulated distributions for $C_{\gamma}$ and $C_{\rm bkg}$
in eq.~\ref{eq_fit}, it is crucial to check
that the $C$ variable is well described in the simulation. 
We show data and MC distributions for photons 
(from radiative lepton pair events and single, isolated photon MC) 
in Fig.~\ref{fig_dmcds}e and clusters from data and MC $\tau$ decays 
($\tau^{\pm}\rightarrow\rho^{\pm}\nu_{\tau}$ and $\rho^{\pm}\rightarrow\pic\ppio$, sec.~4.2 in \cite{taupol}) in
 Fig.~\ref{fig_dmcds}f. The simulation
describes well the $C$ variable distributions
for clusters produced by isolated photons as 
well as by $\ppio$'s. 
In Fig.~\ref{fig_Cvar} we show distributions of the $C$ variable
for background clusters from different sources in the JETSET simulation. 
The shapes of the distributions are similar, although 
$\ppio\rightarrow\gamma\gamma$ and $\eta\rightarrow\gamma\gamma$
decays tend to produce less clusters with higher values of $C$ than 
other sources of background.  

The fitted fraction of photons $f$ is shown in
Tab.~\ref{tab_last} for different cluster energy ranges. 
The comparison of the data and fit results 
is given in Fig.~\ref{fig_Cdatbcg}. 
The contributions from prompt photons and
 background to the fit are shown.  
The $\chi^{2}$, taking into account 
the statistical errors on the $C_{\gamma}$ and $C_{\rm bkg}$ distributions, 
were between 16 and 37 for 23 degrees-of-freedom.

\section{Corrections for efficiency, acceptance and initial state radiation}
\label{sec_cor}

The energy spectrum of  
 prompt photons obtained in the previous section 
was corrected for photons lost in the selection process and outside the 
geometrical acceptance. To
ensure that the energy spectrum of photons was not biased,
 efficiency corrections
were determined separately for each energy bin. Then the contribution due to 
initial state radiation was subtracted 
and the corrected photon energy 
spectrum was normalised to the total number of hadronic events. 

We applied corrections for the following effects. 
\ben
\item Local, temporary inefficiencies in the presampler system.
      This factor was determined from data to be $1.23$ with
       negligible statistical error.

\item Rejection of real photons by cuts 2 and 3. 
      A correction was determined using the data sample of photons in 
      radiative lepton pair events.

\item  Additional rejection, by the combined effects of cuts 1, 2
      and 3, of photons that formed calorimeter clusters with other   
      particles in the event or converted in the beam-pipe or 
      central detector. This correction, called photon environment in
      Tab.~\ref{tab_cor}, was determined with the JETSET Monte 
      Carlo as the ratio of the combined cut efficiency for prompt photons 
      where no other particle contributed to the calorimeter cluster, to
      the efficiency for all prompt photons.   

\item The contribution of initial state radiation, estimated  using 
       the KORALZ program version~4.0,  was subtracted.  
\een
      Values of the energy dependent corrections are shown
      in Tab.~\ref{tab_cor}. 
      To compare our result with theoretical predictions we applied 
      an additional correction for the 
      rejection of photons by the cut on the polar angle 
      $\theta$ ($|\rm{\cos\theta}|<0.72$). The correction was 1.58. We 
      assumed the leading-order 
      $1+\cos^{2}\theta$ dependence of the photon production cross-section,
      which was consistent with the polar angle distribution 
      of the selected calorimeter clusters. 
      The fully corrected energy spectrum of prompt photons in hadronic
      \Zo ~decays is shown in
      Tab.~\ref{tab_last} and in Fig.~\ref{fig_spectrum}. 

\section{Systematic effects}
\label{sec_sys}
   We checked the dependence of our result on 
possible deficiencies in the simulation of the detector and on the particle 
composition of the background. 
The main sources of the systematic uncertainties were estimated as follows:
\ben
\item  The sensitivity of the fit result to the quality of the MC
reproduction
 of the $C$ variable was determined as follows, separately for each 
cluster energy bin.  
The $C_{\gamma}$ and $C_{\rm bkg}$ distributions used in the fit 
were simultaneously scaled by ($1\pm\alpha$), where $\alpha~(\simeq4\%)$ was
the error on the mean value of $C$ from the Monte
Carlo added in quadrature to that from the data.
These modified distributions were then used in the fit. The
differences between these results and those obtained with unmodified
distributions were assigned as the systematic error. These
uncertainties, ranging from 10-23\% and partly correlated between different 
photon energy points, were the dominant source of
systematic error in our measurement.  

\item   The fractions of photons in the data 
obtained from the fit using the background spectrum  predicted by 
the HERWIG MC were in agreement with those obtained using
 the JETSET MC background. 
This showed that, within our statistical precision, the result did 
not depend on the details of the model implementation of the 
parton shower development and hadronisation processes in the MC 
generators.  The difference between the fraction of prompt photons 
obtained from the data fitted with the $C_{\rm bkg}$ 
predicted by the JETSET and HERWIG was assigned as a systematic error. 
\item  Although we relied on the data as much as possible in the 
determination of the efficiency corrections we had to resort to MC
to estimate how the efficiencies for photons 
passing the selection cuts were modified by the presence of other
particles in the event (correction 3 in sec.~\ref{sec_cor}). 
The corrections obtained with JETSET and HERWIG models gave results 
consistent within statistical errors despite the differences in the
 modelling of the prompt photon radiation, parton shower and hadronisation 
for the two models. The difference between corrections obtained with the  
JETSET and HERWIG models was assigned as a systematic error.   
\een
In addition several further checks were performed, none of which produced a 
statistically significant, at the one standard deviation level, 
difference from the result of the default procedure and were 
not included in the systematic error. 
\bit

\item In equation~\ref{eq_fit} the amount of the background in the sample was the fit parameter. 
Therefore, the fitted value of $f$ was not, to first order, sensitive to the
background flux as incorporated in the MC generator. 
In principle, some sensitivity 
to the relative fluxes of different background sources remained, since 
it could change the shape of the $C$ variable distribution of the background.
This sensitivity was estimated by 
repeating the fit with the number of clusters in the background 
produced by $\rm \eta$ mesons,
the second largest source of background, adjusted by
the uncertainty on the $\rm \eta$ yield of $\pm20\%$~\cite{ALEPHeta}.
This resulted in 1.5\% change of the prompt photon yield. 

\item  The systematic error from the modelling of the $C$ variable
      (point~1) was consistent with an 
         estimate which assumed a dependence of the factor $\alpha$
         on the value of $C$ variable. For $C$ greater than 2 the 
          factor $\alpha$
          was put to zero. For lower $C$ values
          a linear dependence was assumed, such that 
         the mean value of $\alpha$ for $C$ between 0 and 2 
          was equal to the value of $\alpha$ used in the default 
         procedure.
\item The fit of equation~\ref{eq_fit} was repeated with 
      the distribution of $C$   
      for prompt photons in JETSET used as $C_{\gamma}$ instead of the default 
      of isolated photons.
\item The values of the cuts  
$\Dcut$, $\Scut$ and for the association cut were changed by $\pm 20~\%$.

\item  The fit result (section~\ref{sec_fdat}) did not depend on 
 the bin size of the fitted distribution 
(changed by a factor of 2) or the upper fit boundary 
(moved between 5 and 10 in $C$).

\item The effect of the 
energy resolution was negligible for the size of the photon energy bins
used. 
The absolute energy scale for the electromagnetic calorimeters
is very well calibrated using Bhabha scattering events
and cross-checked with, for example, the $\ppio$ mass peak position. 

\item    The background of calorimeter clusters from 
$\rm {\tau^{+}\tau^{-}}$ events in the selected sample 
(estimated with KORALZ~4.0 \cite{KORALZ}) was less than 
 $0.5\%$ of the observed photon signal.  
\item   The contamination from clusters produced by the LEP accelerator 
background or cosmic rays, estimated with the events collected using a 
random beam-crossing trigger,  was below the level of 2 clusters per million 
events and no cluster passed the selection criteria.
\eit

  To summarise, the principal systematic error comes from the uncertainty
 on the fitted fraction of photons in the data 
 due to the uncertainty on the quality of the MC reproduction of the 
$C$ variable (point~1). 
 The uncertainties due to the possible dependence of our result on the 
modelling  of the prompt 
 photon radiation, parton shower and hadronisation in the MC generators 
were estimated in points~2 and~3.  
The total systematic uncertainty was calculated as the sum in quadrature
of the propagated errors listed in points 1, 2 and 3, 
as shown in Tab.~\ref{tab_sys}. 
The systematic errors were partly correlated between the different 
photon energy ranges. 
\section{Summary and discussion} 
\label{sec_result}

   The energy spectrum of prompt photons in hadronic \Zo ~decays, corrected 
for the geometrical acceptance as discussed in section \ref{sec_cor}, 
is shown in Tab.~\ref{tab_last} 
and in Fig.~\ref{fig_spectrum}. The inner error bars are statistical,
and the outer combined statistical and systematic 
(added in quadrature). The breakdown of the systematic 
uncertainties is given in  Tab.~\ref{tab_sys}.  

To compare our result with leading-order theoretical predictions for 
the quark-to-photon fragmentation function we use 
the leading-order cross-section for prompt photon production in
\ee ~annihilation given by formula~(12) in \cite{Kunszt}:
\begin{equation} 
\frac{1}{\sigma_{\rm had}}\frac{d\sigma(E_{\gamma})}{dE_{\gamma}}=
       \frac{4}{\sqrt{s}}\sum_{\rm q}w_{\rm q}\Dgq,     
\label{eq_sigma1}
\end{equation}
\noindent where $E_{\gamma}$ is the photon energy;  
         $\sqrt{s}$ is the \ee ~centre of mass energy;  
         $w_{\rm q}$ is the relative contribution from quark flavour q   
($w_{\rm q}={\rm \Gamma_{Z\rightarrow q\overline{q}}}/{\rm \Gamma_{Z\rightarrow {hadrons}}}$); $\sigma_{\rm had}$ is the production cross-section for
\ee ~hadronic events,
and $\Dgq$ is the quark q to photon fragmentation function.
In Fig.~\ref{fig_spectrum} we plot the QCD prediction using an asymptotic 
leading-order $\Dgq$ \cite{EWitten,CHL-L} 
as parametrised by Duke and Owens \cite{DUKEOWENS} with 
$\Q2=M_{Z}^2$ and $\Lambda=0.2~\GeV$.  
In Fig.~\ref{fig_spectrum} 
we also plot the QCD predictions using $\Dgq$ from leading-order (LO)
calculations  by Gl$\rm\ddot{u}$ck, Reya and
Vogt \cite{GRV}. We also show the higher-order (HO) 
prediction for the prompt  
photon production cross-section by Gl$\rm\ddot{u}$ck, Reya and
Vogt \cite{GRV} and the beyond leading logarithm (BLL) 
prediction by Bourhis, Fontannaz and Guillet~\cite{Fontannaz}.
The factorisation schemes used 
were $\rm DIS_{\gamma}$ and \ov{MS} respectively.  
The authors of \cite{GRV} and \cite{Fontannaz} 
included in their calculations non-perturbative effects  
through the vector-meson dominance ansatz, although using different 
experimental inputs.

    Our data are in agreement with these theoretical predictions. The 
experimental precision is not sufficient to discriminate between them. 
The ALEPH data on the production of jets 
containing a photon carrying a substantial 
fraction (above 70\%) of the 
jet energy (Fig.~4 in \cite{ALEPH_frfun}) 
show clear disagreement with 
the Duke-Owens parametrisation~\cite{DUKEOWENS}. 
It was noted in~\cite{Fontannaz} that a possible reason 
is that the ALEPH measurement 
is restricted to only part of the phase-space for photon production in 
hadronic \Zo ~decays.

  To summarise, we have measured the  
inclusive production of prompt photons with energy above 10~\GeV~
in hadronic \Zo ~decays. 
Good agreement is found 
with current QCD predictions for the quark-to-photon 
fragmentation function.

\section*{Acknowledgements}

  We are grateful to Zoltan Kunszt, Werner Vogelsang, Richard Roberts 
and  Nigel Glover for stimulating discussions on the 
theoretical aspects of our measurement. 
We also thank  
M.~Gl$\rm\ddot{u}$ck, E.~Reya, and A.~Vogt, as well as 
L.~Bourhis, M.~Fontannaz, and J.Ph.~Guillet
for providing the numerical results of their calculations.    

We particularly wish to thank the SL Division for the efficient operation
of the LEP accelerator at all energies
 and for
their continuing close cooperation with
our experimental group.  We thank our colleagues from CEA, DAPNIA/SPP,
CE-Saclay for their efforts over the years on the time-of-flight and trigger
systems which we continue to use.  In addition to the support staff at our own
institutions we are pleased to acknowledge the  \\
Department of Energy, USA, \\
National Science Foundation, USA, \\
Particle Physics and Astronomy Research Council, UK, \\
Natural Sciences and Engineering Research Council, Canada, \\
Israel Science Foundation, administered by the Israel
Academy of Science and Humanities, \\
Minerva Gesellschaft, \\
Benoziyo Center for High Energy Physics,\\
Japanese Ministry of Education, Science and Culture (the
Monbusho) and a grant under the Monbusho International
Science Research Program,\\
German Israeli Bi-national Science Foundation (GIF), \\
Bundesministerium f\"ur Bildung, Wissenschaft,
Forschung und Technologie, Germany, \\
National Research Council of Canada, \\
Hungarian Foundation for Scientific Research, OTKA T-016660, 
T023793 and OTKA F-023259.\\

\clearpage

\normalsize

\begin{table}[p]
\caption{Energy dependent cuts and corrections to the photon 
energy spectrum. Corrections for photon efficiency and environment are 
multiplicative, ISR is subtracted.}  
\label{tab_cor}
\begin{center}
\begin{tabular}{|c | c | c || c | c || c |}
\hline     
\Rule $E$, \GeV & \parbox{1.3cm}{$\Dcut$ \\ mrad}  & $\Scut$ & \parbox{3.cm}{isolated photon  
efficiency }
& \parbox{3.cm}{photon \\ environment} & \parbox{3.cm}{ $N_{\gamma}^{ISR}$/GeV/$10^{6}$ events} \\ \hline
\Rule $10-15~~$ & 3. & 1.5 & $2.51\pm0.13$ & $1.34\pm0.10    $ & $6.71\pm0.13$   \\ \hline
\Rule $15-20~~$ & 4. & 2.5 & $2.06\pm0.12$ & $1.36\pm0.11    $ & $2.62\pm0.08$   \\ \hline
\Rule $20-25~~$ & 5. & 4.~ & $1.51\pm0.07$ & $1.37\pm0.10    $ & $1.52\pm0.06$   \\ \hline
\Rule $25-30~~$ & 5. & 5.~ & $1.47\pm0.08$ & $1.25\pm0.09    $ & $1.27\pm0.06$   \\ \hline
\Rule $30-35~~$ & 5. & 5.~ & $1.42\pm0.08$ & $1.20\pm0.09    $ & $1.28\pm0.06$   \\ \hline
\Rule $35-40~~$ & 5. & 5.~ & $1.60\pm0.09$ & $1.14\pm0.10    $ & $1.20\pm0.06$   \\ \hline
\Rule $40-45.6$ & 5. & 5.~ & $1.57\pm0.05$ & $1.03\pm0.13    $ & $2.87\pm0.09$   \\ \hline
\end{tabular}
\end{center}
\end{table}
\normalsize

\begin{table}[p]
\caption{The fraction $f$ of prompt photons
in the selected sample and the corrected
energy spectrum of prompt photons in hadronic \Zo ~events 
(the first error is statistical, the second systematic).}
\label{tab_last}
\begin{center}
\begin{tabular}{| c | c | c | c  c  c |}
\hline     
\Rule $E$, \GeV & Mean $E$, \GeV & $f$, \%  & \multicolumn{3}{c |}{$N_{\gamma}$/GeV/$10^{6}$ events}
\\ \hline  
\Rule $10-15~~$ & 12.0 & $32.7\pm~2.1$ &625  & $\pm  70  $&$\pm99$  \\ \hline
\Rule $15-20~~$ & 17.3 & $18.2\pm~1.7$ &266  & $\pm  36  $&$\pm63$  \\ \hline
\Rule $20-25~~$ & 22.3 & $12.9\pm~1.4$ &142  & $\pm  20  $&$\pm37$  \\ \hline
\Rule $25-30~~$ & 27.3 & $27.9\pm~1.9$ &212  & $\pm  24  $&$\pm39$  \\ \hline
\Rule $30-35~~$ & 32.3 & $25.7\pm~2.8$ &118  & $\pm  17  $&$\pm28$ \\ \hline
\Rule $35-40~~$ & 37.1 & $29.9\pm~5.0$ &~75  & $\pm  15  $&$\pm20$  \\ \hline
\Rule $40-45.6$ & 42.1 & $80.4\pm10.2$ &~48  & $\pm  ~9  $&$\pm~8$  \\ \hline
\end{tabular}
\end{center}
\end{table}
\normalsize

\begin{sidewaystable}[p]
\caption{Contributions to the systematic  error  
on the photon energy spectrum shown in Tab.~\ref{tab_last}.} 
\label{tab_sys}
\begin{center}
\begin{tabular}{| c | c || c | c | c |}
\hline     
\Rule Mean $E$, \GeV & $N_{\gamma}$/GeV/$10^{6}$ events  & \multicolumn{3}{c |}
{Systematic errors on $N_\gamma$/GeV/$10^{6}$ events} \\
\cline{3-5} \Rule
\Rule 
 & & \parbox{3cm}{description of \\ $C$ variable} &
\parbox{3.5cm}{model dependence \\ of background \\ distribution} &
\parbox{3.5cm}{photon acceptance} \\ \hline  
\Rule 12.0  &$625\pm70\pm99$&$\pm60$&$\pm73$&$\pm29$ \\ \hline
\Rule 17.3  &$266\pm36\pm63$&$\pm54$&$\pm31$&$\pm12$ \\ \hline
\Rule 22.3  &$142\pm20\pm37$&$\pm32$&$\pm17$&$\pm~6$ \\ \hline
\Rule 27.3  &$212\pm24\pm39$&$\pm28$&$\pm25$&$\pm10$ \\ \hline
\Rule 32.3  &$118\pm17\pm28$&$\pm23$&$\pm14$&$\pm~6$ \\ \hline
\Rule 37.1  &$~75\pm15\pm20$&$\pm17$&$\pm~9$&$\pm~4$ \\ \hline
\Rule 42.1  &$~48\pm~9\pm~8$&$\pm~5$&$\pm~6$&$\pm~3$ \\ \hline
\end{tabular}
\end{center}
\end{sidewaystable}
\normalsize


\newpage
\begin{figure*}
\begin{minipage}{\minisize}
\bce
\mbox{
\epsfig{file=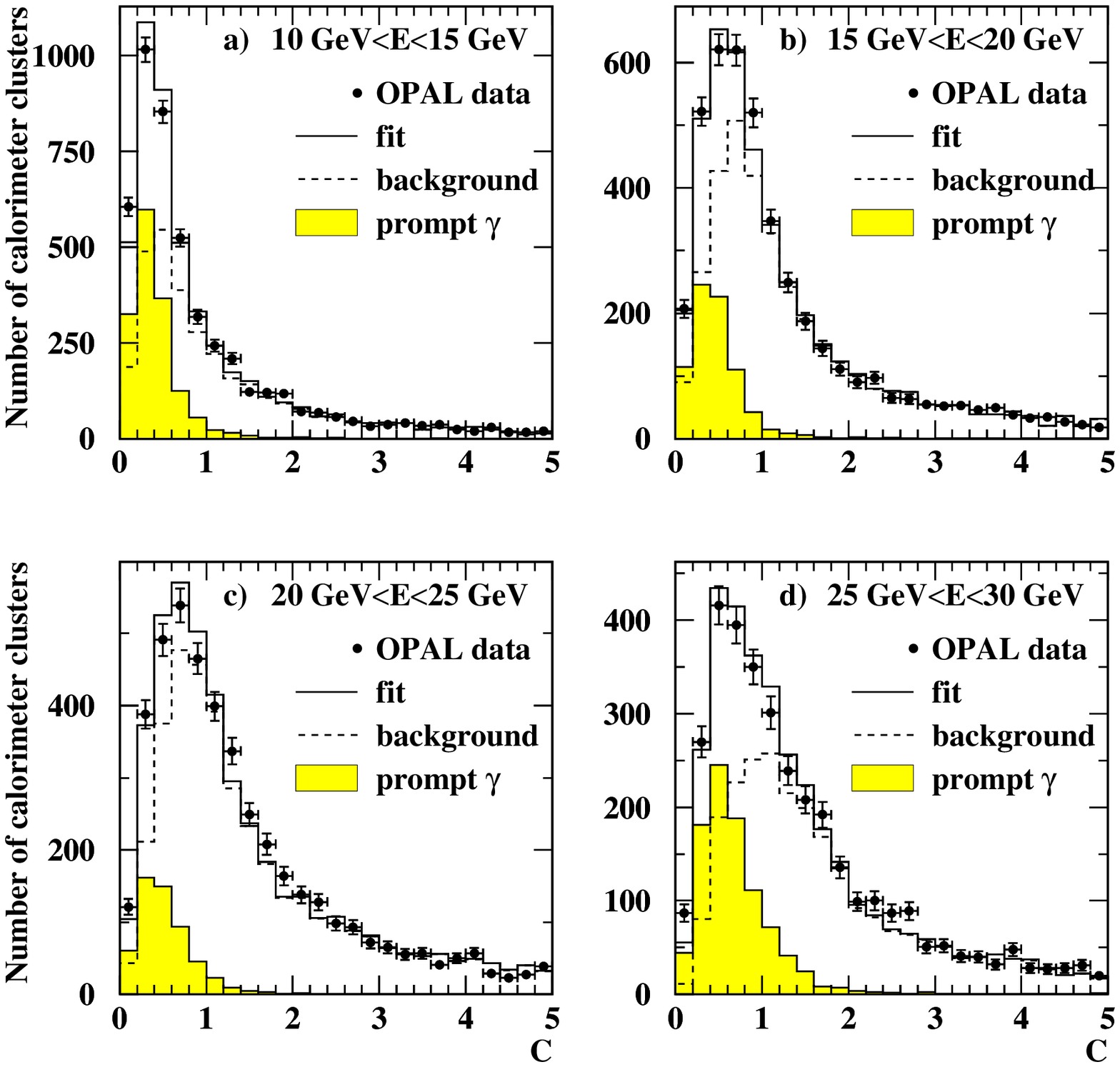,width=\minisize}
}
\ece
\end{minipage}
\end{figure*}

\begin{figure}[p]
\begin{minipage}{\minisize}
\bce
\mbox{
\epsfig{file=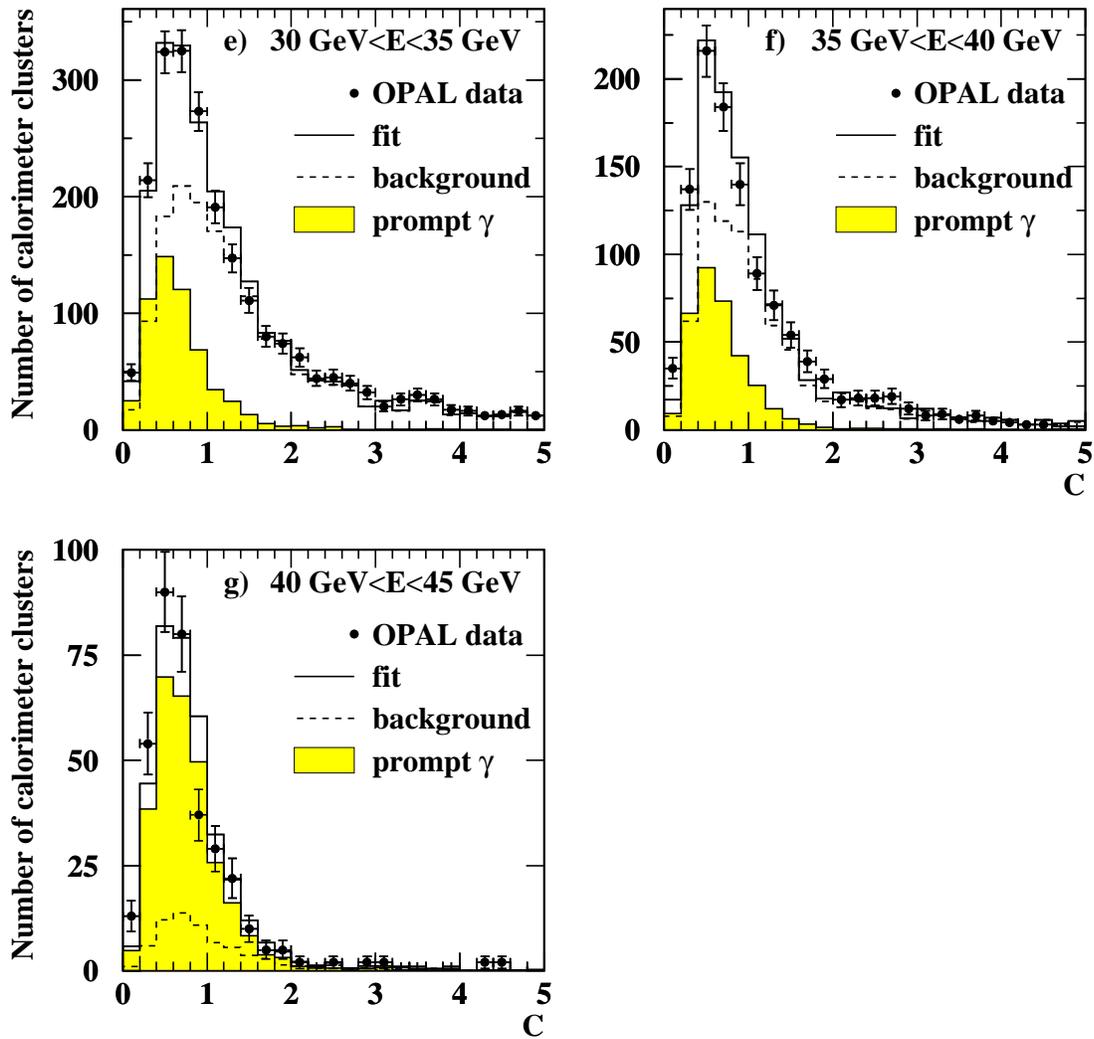,width=\minisize}
}
\caption{Comparison of the data and fit for different 
cluster energy ranges. Contributions from prompt photons and background in the
fit are shown.} 
\label{fig_Cdatbcg}
\ece
\end{minipage}
\end{figure}

\begin{figure}[p]
\begin{minipage}{\minisize}
\bce
\mbox{
\epsfig{file=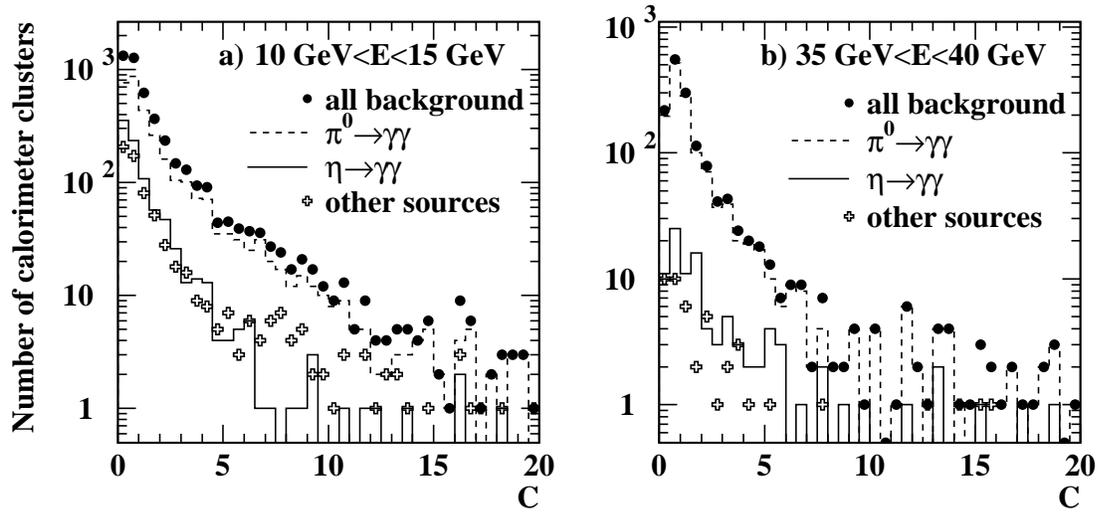,width=\minisize}
}
\vspace{-7.5cm}
\caption{$C$ variable distributions for background clusters from different  
sources in the JETSET simulation for different cluster energy ranges.}
\label{fig_Cvar}
\ece
\end{minipage}
\end{figure}

\begin{figure}[p]
\begin{minipage}{\minisize}
\bce
\mbox{
\epsfig{file=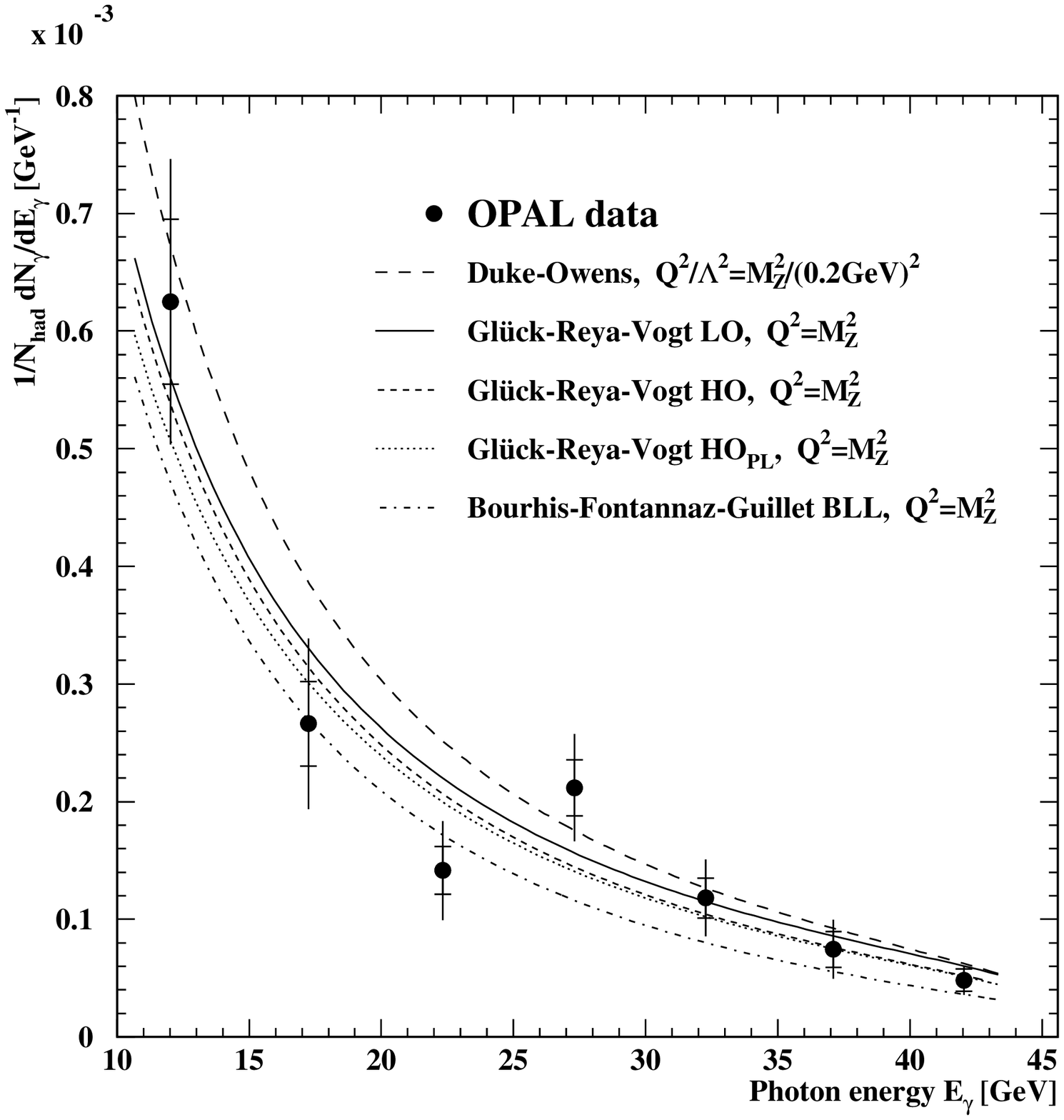,width=\minisize}
}
\caption{The photon energy spectrum in hadronic 
\Zo ~decays compared to various theoretical predictions:  
the Duke-Owens parametrisation~[8], 
the Gl$\rm\ddot{u}$ck, Reya and Vogt predictions including 
leading-order (LO), higher-order (HO) and 
higher-order without the non-perturbative corrections
($\rm HO_{PL}$)~[9]. 
The Bourhis, Fontannaz and Guillet prediction shown include effects 
beyond leading 
logarithms (BLL)~[10]. 
}
\label{fig_spectrum}
\ece
\end{minipage}
\end{figure}


\begin{thebibliography}{99}
\bibitem{CELLO}
CELLO Collab., H.J.~Behrend et al., Z.Phys.C14 (1982)
 189,\\
CELLO Collab., H.J.~Behrend et al., Z.Phys.C20 (1983) 207.  
\bibitem{JADE}
JADE Collab., D.D.~Pitzl et al., Z.Phys.C46 (1990) 1, \\ 
(err.ibid. 47 (1990) 676), \\
JADE Collab., W.~Bartel et al., Z.Phys.C28 (1985) 343. 
\bibitem{MAC}
MAC Collab., E.~Fernandez et al., Phys.~Rev.~Lett. 54 (1985) 95. 
\bibitem{TASSO}
TASSO Collab., W.~Braunschweig et al., Z.Phys.C41 (1988) 385.
\bibitem{VENUS}
VENUS Collab., K.~Abe et al., Phys.~Rev.~Lett. 63 (1989) 1776. 
\bibitem{EWitten}
E.~Witten, Nucl.~Phys. B120 (1977) 189.  
\bibitem{CHL-L}
C.H.~Llewellyn~Smith, Phys.~Lett. B79 (1978) 83.
\bibitem{DUKEOWENS}
J.F.~Owens, Reviews of Modern Physics 59 (1987) 465, \\
D.W.~Duke, J.F.~Owens, Phys.~Rev.~D26 (1982) 1600. 
\bibitem{GRV}
M.~Gl$\rm\ddot{u}$ck, E.~Reya, A.~Vogt, Phys.~Rev.~D48 (1993) 116. 
\bibitem{Fontannaz}
L.~Bourhis, M.~Fontannaz, J.Ph.~Guillet, preprint hep-ph~9704447.
\bibitem{OPAL1}
OPAL Collab., M.Z.~Akrawy et al., Phys.~Lett.~B246 (1990) 285.  

\bibitem{Peter}
E.~Laermann, T.F.~Walsh, I.~Schmitt, P.M.~Zerwas, Nucl.~Phys. B207 (1982) 205
P.~M$\rm{\ddot{a}}$ttig, W.~Zeuner, Z.Phys.C52 (1991) 37.

\bibitem{ALEPH}
ALEPH Collab., D.~Decamp et al., Phys.~Lett. B264 (1991) 476, \\
ALEPH Collab., D.~Buskulic et al., Zeit. Phys.C57 (1993) 17. 
\bibitem{DELPHI}
DELPHI Collab., P.~Abreu et al., Zeit. Phys.C53 (1992) 555, \\
DELPHI Collab., P.~Abreu et al., Zeit. Phys.C69 (1995) 1.
\bibitem{L3}
L3 Collab., O.~Adriani et al., Phys.~Lett. B292 (1992) 472, \\ 
L3 Collab., O.~Adriani et al., Phys.~Lett. B297 (1993) 469, \\
L3 Collab., O.~Adriani et al., Phys.~Lett. B301 (1993) 136, \\
L3 Collab., M.~Acciarri et al., Phys.~Lett. B388 (1996) 409. 
\bibitem{OPAL}
OPAL Collab., G.~Alexander et al., Phys.~Lett. B264 (1991) 219, \\
OPAL Collab., P.D.~Acton et al., Zeit. Phys.C54 (1992) 193, \\    
OPAL Collab., P.D.~Acton et al., Zeit. Phys.C58 (1993) 405, \\
OPAL Collab., R.~Akers et al., Zeit. Phys.C67 (1995) 15, \\ 
OPAL Collab., R.~Akers et al., Zeit. Phys.C68 (1995) 531, \\ 
OPAL Collab., G.~Alexander et al.,  Zeit. Phys.C71 (1996) 1.  
\bibitem{Glover}
E.W.N.~Glover, A.G.~Morgan, Z.Phys.C62 (1994) 311.  
\bibitem{ALEPH_frfun}
ALEPH Collab.,D.~Buskulic et al., Z.Phys.C69 (1996) 365.
\bibitem{Kunszt}
Z.~Kunszt, Z.~Tr\'ocs\'anyi, Nucl. Phys. B394 (1993) 139.
\bibitem{OPALNIM}
OPAL Collab., K.~Ahmet et al., Nucl. Instrum. Methods A~305~(1991)~275.
\bibitem{Hadsel_paper}
OPAL Collab., G.~Alexander et al., Z.Phys.C52 (1991) 175.
\bibitem{JETSET}
T.~Sj{\"o}strand, Comput. Phys.~Commun. 39 (1986) 347, \\
T.~Sj{\"o}strand, M.Bengtsson, Comput.~Phys.~Commun. 43 (1987) 43.
\bibitem{HERWIG}
G.~Marchesini et~al., Comput.~Phys.~Commun. 67 (1992) 465.
\bibitem{JETSETUNE}
OPAL Collab., G.~Alexander et al., Z.Phys.C69 (1996) 543.
\bibitem{GOPAL}
J.~Allison et~al., Nucl. Instrum. Methods A 317 (1992) 47.
\bibitem{taupol}
OPAL Collab., R.~Akers et al., Phys.~Lett. B328 (1994) 207. 
\bibitem{RBarlow}
R.~Barlow, J.~Comp.~Phys. 72 (1987) 202, \\
R.~Barlow, C~.Beeston, Comput.~Phys.~Commun. 77 (1993) 219. 
\bibitem{KORALZ}
Z.~Jadach, B.~Ward, Z.~Was, Comput.~Phys.~Commun. 79 (1994) 503.  
\bibitem{ALEPHeta}
ALEPH Collab., D.~Buskulic et al., Phys.~Lett. B292 (1992) 210.



\end{thebibliography}
\end{document}